\documentclass[11pt]{article}  % Leave intact
\usepackage{adassconf}
\usepackage{hyperref}

\begin{document}   % Leave intact

%-----------------------------------------------------------------------
%			    Paper ID Code
%-----------------------------------------------------------------------
% Enter the proper paper identification code.  The ID code for your
% paper is the session number associated with your presentation as
% published in the official ADASS 98 conference proceedings.  You can
% find this number locating your abstract in the printed proceedings
% that you received at the meeting or on-line via
% http://monet.astro.uiuc.edu/adass98/detailedprog.html; the ID code
% is the letter-number sequence proceeding the title of your
% presentation.  
%
% This will not appear in your paper; however, it allows different
% papers in the proceedings to cross-reference each other.
%
% EXAMPLE: \paperID{T1.9}
% EXAMPLE: \paperID{D21}
% EXAMPLE: \paperID{P10.4}
%
% Note that you should only have one \paperID, and it should not
% include a trailing period.  

\paperID{T3.2}

%-----------------------------------------------------------------------
%		            Paper Title 
%-----------------------------------------------------------------------
% Enter the title of the paper.
%
% EXAMPLE: \title{A Breakthrough in Astronomical Software Development}
% 

\title{The Pixon Method of Image Reconstruction}

%-----------------------------------------------------------------------
%		          Authors of Paper
%-----------------------------------------------------------------------
% Enter the authors followed by their affiliations.  The \author and
% \affil commands may appear multiple times as necessary (see example
% below).  List each author by giving the first name or initials first
% followed by the last name.  Authors with the same affiliations
% should grouped together. 
%
% EXAMPLE: \author{Raymond Plante, Doug Roberts, 
%                  R.\ M.\ Crutcher\altaffilmark{1}}
%          \affil{National Center for Supercomputing Applications, 
%                 University of Illinois Urbana-Champaign, Urbana, IL
%                 61801}
%          \author{Tom Troland}
%          \affil{University of Kentucky}
%
%          \altaffiltext{1}{Astronomy Department, UIUC}
%
% In this example, the first three authors, "Plante", "Roberts", and
% "Crutcher" are affiliated with "NCSA".  "Crutcher" has an alternate 
% affiliation with the "Astronomy Department".  The fourth author,
% "Troland", is affiliated with "University of Kentucky"

\author{Richard C. Puetter}
\affil{Center for Astrophysics and Space Sciences, University of California,
San Diego, La Jolla, CA 92093-0424, USA}
\author{Amos Yahil} 
\affil{Department of Physics and Astronomy, State University of New York, Stony
Brook, NY 11794-3800, USA}

%-----------------------------------------------------------------------
%			 Contact Information
%-----------------------------------------------------------------------
% This information will not appear in the paper but will be used by
% the editors in case you need to be contacted concerning your
% submission.  Enter your name as the contact along with your email
% address.
% 
% EXAMPLE:  \contact{Raymond Plante}
%           \email{rplante@ncsa.uiuc.edu}
%

\contact{Rick Puetter}
\email{rpuetter@ucsd.edu}

%-----------------------------------------------------------------------
%		      Author Index Specification
%-----------------------------------------------------------------------
% Specify how each author name should appear in the author index.  The 
% \paindex{ } should be used to indicate the primary author, and the
% \aindex for all other co-authors.  You MUST use the following
% syntax: 
%
% SYNTAX:  \aindex{LASTNAME, F. M.}
% 
% where F is the first initial and M is the second initial (if
% used).  This guarantees that authors that appear in multiple papers
% will appear only once in the author index.  
%
% EXAMPLE: \paindex{Plante, R. L.}
%          \aindex{Roberts, D. A.}
%          \aindex{Crutcher, R. M.}
%          \aindex{Troland, T.}

\paindex{Puetter, R. C.}
\aindex{Yahil, A.}     % Remove this line if there is only one author

%-----------------------------------------------------------------------
%			Subject Index keywords
%-----------------------------------------------------------------------
% Enter up to 6 keywords describing your paper.  These will NOT be
% printed as part of your paper; however, they will be used to
% generate the subject index for the proceedings.  There is no
% standard list; however, you can consult the indices for past ADASS
% proceedings (http://iraf.noao.edu/ADASS/adass.html). 
%
% EXAMPLE:  \keywords{visualization, astronomy: radio, parallel
%                     computing, AIPS++, Galactic Center}
%
% In this example, the author noticed that "radio astronomy" appeared
% in the ADASS VII Index as "astronomy" being the major keyword and
% "radio" as the minor keyword.

\keywords{Pixon method, image reconstruction, image processing,
multiresolution, inverse problem}

%-----------------------------------------------------------------------
%			       Abstract
%-----------------------------------------------------------------------
% Type abstract in the space below.  Consult the 
% (http://monet.astro.uiuc.edu/adass98/proceedings/LatexSummary.ps)
% for a list of supported macros (e.g. for typesetting special
% symbols). 

\begin{abstract}          % Leave intact
% Place the text of your abstract here 

The Pixon method is a high-performance, nonlinear image reconstruction method
that provides statistically unbiased photometry and robust rejection of
spurious sources.  Relative to other methods, it can increase linear spatial
resolution by a factor of a few and sensitivity by an order of magnitude or
more.  All of these benefits are achieved in computation times that can be
orders of magnitude faster than its best competitors.

\end{abstract}

%-----------------------------------------------------------------------
%			      Main Body
%-----------------------------------------------------------------------
% Place the text for the main body of the paper here.  You should use
% the \section command to label the various sections; use of
% \subsection is optional.  Significant words in section titles should
% be capitalized.  Sections and subsections will be numbered
% automatically. 
%
% EXAMPLE:  \section{Introduction}
%           ...
%           \subsection{Our View of the World}
%           ...
%           \section{A New Approach}
%
% It is recommended that you look at the sample papers, sample1.tex
% and sample2.tex, for examples for formatting references, footnotes,
% figures, equations, html links, lists, and other special features.  

\section{Introduction}

Optimal extraction of the underlying quantities from measured data requires the
removal of measurement defects such as noise and limited instrumental
resolution.  When the underlying quantity is an image, this process is known as
image reconstruction (sometimes called image restoration, if the data are also
in the form of an image).

The original Pixon method (Pi\~na \& Puetter 1993; Puetter \& Pi\~na 1993;
Puetter 1995, 1997) was developed to eliminate problems with existing image
reconstruction techniques, particularly signal-correlated residuals and the
production of spurious sources.  This was followed by the accelerated Pixon
method that vastly increased the computational speed (Yahil and Puetter 1995,
unpublished).  Recently, a quick Pixon method was developed that is even
faster, at the expense of some photometric inaccuracy for low
signal-to-noise-ratio features (Puetter and Yahil 1998, unpublished).
Nevertheless, the quick Pixon method provides excellent results for a wide
range of imagery and, with special-purpose hardware now under design, is
capable of real-time video image reconstruction.

Since its inception, the Pixon method in its various forms has been applied to
a wide variety of astronomical, surveillance, and medical image reconstructions
(spanning all wavelengths from $\gamma$-rays to radio), as well as to
spectroscopic data.  In all cases tested so far, both by the authors and by a
variety of other workers, the Pixon method has proved superior in quality to
all other methods and computationally much faster than its best competitors.

A patent for the Pixon method is pending, restricting its commercial use.  For
individual scientific purposes, the original Pixon method is freely available
in IDL and C++ from the \htmladdnormallink{San Diego Supercomputer Center}
{http://pixon.sdsc.edu}.  The accelerated and quick Pixon methods are sold by
our company, \htmladdnormallink{Pixon LLC} {http://www.pixon.com}, but special
arrangements for their free use in selected scientific projects can be made.

The next sections discuss image reconstruction in general
(\S\ref{sect:reconstruction}), describe the Pixon method (\S\ref{sect:Pixon}),
and give some examples (\S\ref{sect:examples}).  The HTML and PDF versions of
this paper contain numerous additional examples via clickable hyperlinks.  The
complete set of public image reconstructions, including
\htmladdnormallink{videos} {http://www.pixon.com/brochure.html\#Movies}, can be
seen on our Web site at \htmladdnormallink{http://www.pixon.com}
{http://www.pixon.com}.

\section{Image Reconstruction Methods\label{sect:reconstruction}}

For data taken with linear detectors, image reconstruction often becomes a
matter of inverting an integral relation of the form
\begin{equation}
D({\bf x}) = \int {\bf dy} H({\bf x},{\bf y}) I({\bf y}) + N({\bf x})
\quad. \label{eq:convolve}
\end{equation}
Here $D$ are the data, $I$ is the sought underlying image model, $H$ is the
point-spread function (PSF) due to instrumental and possibly atmospheric
blurring, and $N$ is the noise.  To avoid confusion, we use the term image to
refer exclusively to the true underlying image or its model.  Contrary to
common parlance, the data are never called the image.  We clearly distinguish
between abstract image space and real data space, with the PSF transforming
from the former to the latter.

If the PSF is only a function of the displacement, $H({\bf x},{\bf y}) = H({\bf
x - y})$, then the integral in equation (\ref{eq:convolve}) becomes a
convolution, but in general the PSF can vary across the field.  Note that the
data need not have the same resolution as the image, or even the same
dimensionality.  For example, the Infrared Astronomical Satellite (IRAS)
provided 1--D scans across the 2--D sky, and tomography data consist of
multiple 2--D projections of 3--D images.  Another common case is of data
consisting of multiple, dithered, exposures of the same field, which all need
to be modeled by a single image, possibly with PSFs that vary from one exposure
to another.

Image reconstruction differs from standard solutions of integral equations due
to the noise term, $N$, whose nature is only known statistically.  Methods for
solving such an equation fall under two broad categories. (1) Direct methods
apply explicit operators to the data to provide estimates of the image.  These
methods are often linear, or have very simple nonlinear components.  Their
advantage is speed, but they typically amplify noise, particularly at high
frequencies.  (2) By contrast, indirect methods model the noiseless image,
transform it only forward to provide a noise-free data model, and fit the
parameters of the image to minimize the residuals between the real data and the
noise-free data model.  The advantage of indirect methods is that noise is
supposedly excluded (but see \S\ref{sect:NNLS}).  Their disadvantage is the
required modeling of the image.  If a good parametric form for the image is
known {\em a priori\/}, the result can be superb.  If not, either the derived
image badly fits the data or, conversely, it overfits the data, interpreting
noise as real features. Indirect methods are typically significantly nonlinear
and much slower than direct methods.

The Pixon method is a nonparametric, indirect, reconstruction method.  It
avoids the pitfalls of other indirect methods, while achieving a computational
speed on a par with direct methods.  To appreciate its advantages we therefore
first provide a brief description of some competing direct and indirect
methods.

\subsection{Direct Methods\label{sect:direct}}

In the case of a position-independent PSF, naive image deconvolution is
obtained by inverting the integral equation (\ref{eq:convolve}) in Fourier
space, ignoring the noise:
\begin{equation}
\tilde I({\bf k}) = \tilde D({\bf k})/\tilde H({\bf k}) \quad,
\label{eq:deconvolve}
\end{equation}
where the tilde designates Fourier transform, and ${\bf k}$ is the spatial
wavenumber.  This method amplifies noise at high frequencies, since $H({\bf
k})$ is a rapidly declining function of $|{\bf k}|$, while $D({\bf k})$, which
includes high-frequency noise, is not.

To minimize noise amplification, nonlinear filtering is often applied to the
data prior to Fourier inversion.  Common filters are those due to Wiener (1949;
see also Press et al.\ 1992) and thresholding of wavelet transforms (e.g.,
Donoho \& Johnstone 1994; Donoho 1995).  The Wiener filter has the advantage of
being applied in Fourier space.  Wavelet thresholding requires a wavelet
transform, thresholding, and a back wavelet transform, all followed by Fourier
inversion.  Demonstrations of the superior performance of the Pixon method
relative to Wiener-filtered Fourier inversions can be seen on our
\htmladdnormallink{Web site} {http://www.pixon.com} in the reconstruction of
the \htmladdnormallink{Lena}
{http://www.pixon.com/figures/FIG_Lena_comparison.html} image and an aerial
view of \htmladdnormallink{New York City}
{http://www.pixon.com/figures/FIG_NYC_comparison.html}.

\subsection{Parametric Least-Squares Fit}

The simplest indirect image reconstruction is a least-squares fit of a
parametric model with a small number of parameters compared to the number of
data points.  Originally due to Gauss, this method is always superior to other
methods, provided that the image can be so modeled.  It is equivalent to a
maximum-likelihood optimization in which the residuals are assumed to be
Gaussian distributed.  All models within the restricted parameterization are
considered equal\-ly likely, and the likelihood function maximized is $P(D|I)$,
the conditional probability of the data, given the model image.  For this
reason, the method is also known as maximum {\em a posteriori\/} probability
(MAP) image reconstruction.

\subsection{Nonnegative Least-Squares Fit\label{sect:NNLS}}

When no parametric model of the image is known, the number of image model
parameters can quickly become comparable to, or exceed, the number of data
points. In this case, a MAP solution becomes inappropriate.  For example, if
the number of points in the image model equals the number of data points, then
the nonsingular nature of the linear integral transform in equation
(\ref{eq:convolve}) assures that there is a solution for which the data,
including the noise, are exactly modeled with zero residuals.  This is clearly
the same poor solution, with all its noise amplification, obtained by the naive
Fourier deconvolution.

The above example shows that an unrestricted indirect method is no better at
controlling noise than a direct method, and therefore the model image must be
restricted in some way.  The indirect methods described below all restrict the
image model and differ only in the specifics of image restriction.

A simple restriction is to constrain the model image to be positive.  Since
even a delta-function image is broadened by the PSF, it follows that the exact
inverse of any noisy data with fluctuations on scales smaller than the PSF must
be both positive and negative.  By preventing the image from becoming negative,
the noise-free data model cannot fluctuate on scales smaller than the PSF,
which is equivalent to smoothing the data on the scale of the PSF.

While smoothing over the scale of the PSF helps to reduce noise fitting, it
does not go far enough.  The Pixon method (\S\ref{sect:Pixon}) smoothes further
where possible, and is therefore able to eliminate noise fitting on larger
scales.  Demonstrations of its superior performance relative to nonnegative
least squares can be seen on our \htmladdnormallink{Web site}
{http://www.pixon.com} in the reconstruction of $\gamma$-ray imaging in the
direction of \htmladdnormallink{Virgo}
{http://www.pixon.com/figures/FIG_OSSE_Virgo.html}, as well as in the
\htmladdnormallink{Lena}
{http://www.pixon.com/figures/FIG_Lena_comparison.html} and
\htmladdnormallink{New York City}
{http://www.pixon.com/figures/FIG_NYC_comparison.html} reconstructions.

\subsection{Maximum-Entropy Method}

Bayesian methods go a step further by assigning explicit {\em a priori\/}
probabilities to different models and then maximizing the joint probability of
the model and the data:
\begin{equation}
P(I\cap D) = P(D|I)P(I) = P(I|D)P(D) \quad. \label{eq:Bayes}
\end{equation}
The model probability function, $P(I)$, is known as the prior.  For example,
the nonnegative least-squares method (\S \ref{sect:NNLS}) is a Bayesian method
with all nonnegative models assigned equal nonzero probability and all other
models zero probability.

The rationale behind Bayesian methods is to optimize $P(I|D)$, the conditional
probability of the image given the data.  This probability is proportional to
the joint probability $P(I\cap D)$, since the data are fixed, and $P(D)$ can be
viewed as a constant.  However, it is important to recognize that, unlike
$P(D|I)$ which depends on the instrumental response function and noise
statistics, $P(I|D)$ is not known from first principles.  In practice,
therefore, computing $P(I\cap D)$ requires the specification of a completely
arbitrary prior, $P(I)$.

The maximum-entropy method (MEM) is the most popular Bayesian image
reconstruction technique.  It uses the prior
\begin{equation}
P(I) = \exp\left( - \alpha \sum_i I_i \ln I_i \right) \quad, \label{eq:MEM}
\end{equation}
where the sum is over the image pixels, and $\alpha$ is an adjustable parameter
designed to strike a balance between image smoothness and goodness of fit.
(See Gull 1989 and Skilling 1989 for a discussion of a ``natural'' choice for
$\alpha$.)

  The sum in equation (\ref{eq:MEM}) approximates the information entropy of
Shannon (1948), which is maximized for a homogeneous image.  By favoring a flat
image, MEM therefore eliminates structure not required by the data and
suppresses noise fitting.  The fundamental difficulty with this approach is
that the MEM prior is a global constraint.  (Specifically, the prior is
invariant under random scrambling of the pixels.)  MEM therefore enforces an
average smoothness on the entire image and does not recognize that the density
of information content in the image varies from location to location.  Hence,
MEM must necessarily oversmooth the image in some regions and undersmooth it in
others.  More sophisticated MEM schemes try to remedy this situation by
applying separate MEM priors in different parts of the image, or by modifying
the logarithmic term to $\ln(I_i/M_ie)$, where $M$ is some preassigned model
and $e$ is the base of the natural logarithm (Burch, Gull, \& Skilling 1983),
but there is additional arbitrariness in the choice of $M$.

By contrast, the Pixon method (\S\ref{sect:Pixon}) adapts itself to the
distribution of information content in the image.  Demonstrations of its
superior performance relative to MEM are given below (\S\ref{sect:examples})
and on our \htmladdnormallink{Web site} {http://www.pixon.com} in a
reconstruction of hard X-ray observations of a \htmladdnormallink{solar flare}
{http://www.pixon.com/figures/FIG_YOHKOH.html}.

\section{The Pixon Method\label{sect:Pixon}}

Unlike Bayesian methods, the Pixon method (Pi\~na \& Puetter 1993; Puetter \&
Pi\~na 1993; Puetter 1995, 1997) does not assign explicit prior probabilities
to image models.  Instead, it restricts them by seeking minimum complexity
(Solomonoff 1964; Kolmogorov 1965; Chaitin 1966), which not only enables an
efficient representation of the image but is also the best way to separate
signal from noise.

In simple terms, the Pixon method implements the principle of {\em Ockham's
Razor\/} to select the simplest plausible model of the image.  Clearly, if the
signal in the image can be adequately represented by a minimum of $P$
parameters, adding another parameter only serves to introduce artifacts by
fitting the noise.  Conversely, the removal of a parameter results in an
improper representation of the image, since adequate fits to the image require
a minimum of $P$ parameters.

While few would dispute that a model with minimum complexity (also called
algorithmic information content) is optimal, in practice it is impossible to
find such a model for any but the most trivial problems.  For example, one
might try to model an image as the smallest number of contiguous patches of
homogeneous intensity that still adequately fit the data.  While there clearly
is such a solution, it is quite another matter to find it among the
combinatorially large number of possible patch patterns.  And we have not even
begun to consider patches that are not completely homogeneous.

The Pixon method overcomes this difficulty in the same practical spirit in
which other combinatorial problems have been solved, such as the famous
traveling salesman problem (e.g., Press et al.\ 1992).  One finds an
intelligent scheme in which complexity is reduced significantly in a manageable
number of iterations.  After that, the decline in complexity per iteration
drops sharply, and the process is halted.  The solution found in this manner
may not have the ultimate minimum complexity, but it is already so superior to
other models that it is worth adopting.

The Pixon method minimizes complexity by smoothing the image model {\em
locally\/} as much as the data allow, thus reducing the number of independent
patches, or Pixon elements, in the image.  Formally, the image is written as an
integral over a pseudoimage
\begin{equation}
I({\bf y}) = \int {\bf dz} K({\bf y},{\bf z}) \phi({\bf z}) \quad,
\label{eq:Pixon}
\end{equation}
with a positive kernel function, $K$, designed to provide the smoothing.  As in
the case of the nonnegative least-squares fit, requiring the pseudoimage,
$\phi$, to be positive eliminates fluctuations in the image, $I$, on scales
smaller than the width of $K$.  Importantly, this scale is adapted to the data.
At each location, it is allowed to increase in size as much as possible without
violating the local goodness of fit (GOF).  Where the kernel is a delta
function, the Pixon element spans one pixel, and the Pixon method reduces to a
nonnegative least-squares fit, with the noise-free data model smooth on the
scale of the PSF.  Where the data allow smoothing on larger scales, however,
the Pixon elements become larger.  As a result, the overall number of Pixon
elements in the image drops, and complexity is reduced.

Complexity can be reduced not only by having kernel functions of different
sizes to allow for multiresolution, but also by a judicious choice of their
shapes.  For example, circularly symmetric kernels, which might be adequate for
the reconstruction of most astronomical images, are not the most efficient
smoothing kernels for images with elongated features, e.g., an aerial
photograph of a city.  Altogether the choice of kernels is the language by
which the image model is specified, which should be rich enough to characterize
all the independent elements of the image.  We have found, in practice, that
circular kernels spanning 3--5 octaves in size, with 2--4 kernels per octave,
are adequate for most image reconstructions.  Additional elliptical kernels are
needed only for images with clearly elongated features.

The Pixon reconstruction consists of a simultaneous search for the broadest
possible kernel functions and their associated pseudoimage values that together
provide an adequate fit to the data.  In practice, the details of the search
vary depending on the flavor of the Pixon method used.  Generally, however, one
alternately solves for the pseudoimage given a Pixon map of kernel functions
and then attempts to increase the scale sizes of the kernel functions given the
current image values.  The number of iterations required varies depending on
the complexity of the image, but for most problems a couple of iterations
suffice.

The essence of the Pixon method is the imposition of the local criteria by
which the kernel functions are chosen.  For each pixel in pseudoimage space,
the combination of the Pixon kernel and the PSF define a footprint in data
space.  We accept the largest Pixon kernel whose GOF and signal-to-noise ratio
(SNR) within that footprint pass predetermined acceptance conditions set by the
user.  If no kernel has adequate GOF we assign a delta-function kernel,
provided that the SNR for its footprint is high enough.  If the SNR also fails
to meet our condition, no kernel is assigned.

The strength of the Pixon method is in its rejection of features that do not
meet strict statistical acceptance criteria.  Precisely because of this
conservatism, which results in a significant lowering of the noise floor, the
Pixon reconstruction is able to find weak but significant features missed by
other methods and to resolve all features better.  Sensitivity is often
improved by an order of magnitude or more relative to competing methods and
resolution by a factor of a few.

Note that the SNR required by the Pixon method is not per pixel in the data,
but the overall SNR in the data footprint of the Pixon kernel.  The Pixon
method is just as powerful in detecting large, low-surface-brightness features
as small features with higher surface brightness.  Acceptance or rejection of
the feature is based in both cases on the {\em statistical significance\/}
demanded by the user.  Demonstrations of the ability of the Pixon method to
reconstruct images with low SNR can be seen in the mammogram example (Figure
\ref{fig:mammogram} below) and in the reconstruction of $\gamma$-ray imaging in
the direction of \htmladdnormallink{Virgo}
{http://www.pixon.com/figures/FIG_OSSE_Virgo.html} on our
\htmladdnormallink{Web site} {http://www.pixon.com}.

Finally, note that with a very terse representation that only has a small
number of nonzero pseudoimage values, the image is also represented in a very
compressed form.  In the field of data compression one normally distinguishes
between strong but lossy versus moderate and nonlossy compression.  The Pixon
method defines a new type of noiseless compression.  The signal is preserved
and restorable, while the noise is eliminated.  This mode of compression is
ideal, since compression can easily reach a factor of 100 for a typical image,
yet the only loss is that of unwanted noise.  A compressed form of the code is
now under design; current codes, however, are primarily built for accuracy and
speed and do not yet achieve such high compression.

\begin{figure}
\plotone{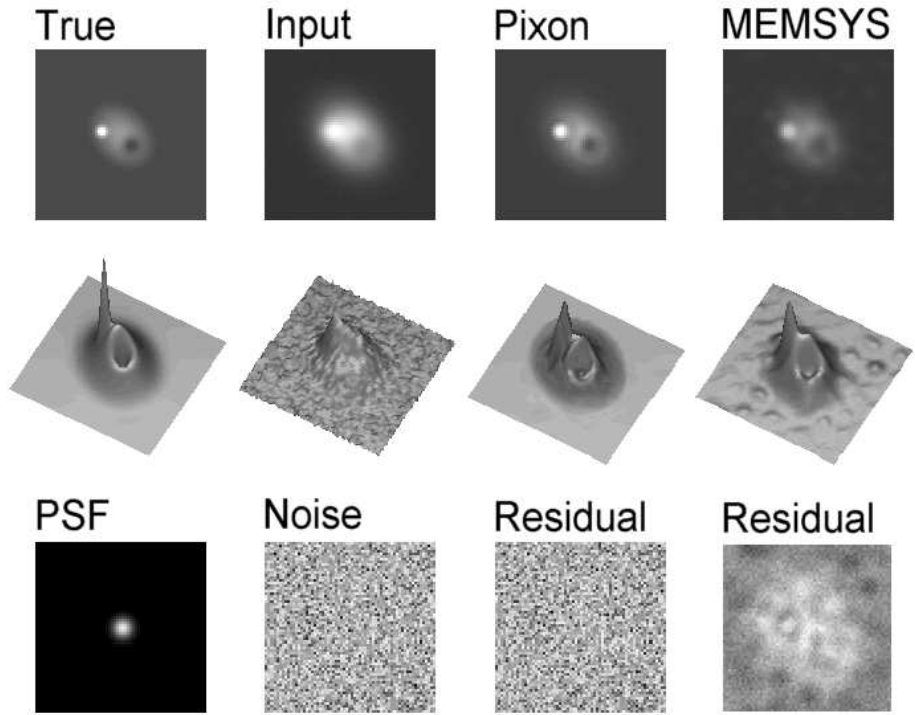}
\caption{Pixon and MEM restorations, showing PSF (Gaussian, FWHM=6 pixels),
true image, Gaussian noise, restored images and residuals.  Peak SNR=20; SNR=12
at the bottom of the ``hole''.  (Reproduced from Puetter 1994.)
\label{fig:synthetic}}
\end{figure}

\section{Examples of Pixon Reconstructions\label{sect:examples}}

The Pixon method has now been used in a variety of applications by the authors
and others (see, Puetter 1995, 1997 and references therein).  In this section
we present a few examples of Pixon reconstructions for a variety of
applications.  Additional examples can be found in the
\htmladdnormallink{brochure} {http://www.pixon.com/brochure.html} and
\htmladdnormallink{figures}{http://www.pixon.com/figures} on our
\htmladdnormallink{Web site} {http://www.pixon.com}.

\subsection{A Synthetic Data Set}

The example presented in Figure \ref{fig:synthetic} (or its
\htmladdnormallink{color version}
{http://www.pixon.com/figures/FIG_p_and_h.html} on our \htmladdnormallink{Web
site} {http://www.pixon.com}) provides one of the best examples of comparative
results of MEM and the Pixon method for a synthetic data set, i.e. a data set
with known properties.  The MEM algorithm used was MEMSYS 5, the most current
release of the MEMSYS algorithms developed by Gull and Skilling (Gull \&
Skilling 1991).  The MEMSYS reconstruction was performed by Nick Weir, and was
enhanced by his multichannel correlation method (Weir 1991).  As can be seen,
the Pixon reconstruction is superior to the multichannel MEMSYS result, and is
free of the low-level spurious sources and signal-correlated residuals evident
in the MEMSYS reconstruction.

\begin{figure}
\plotone{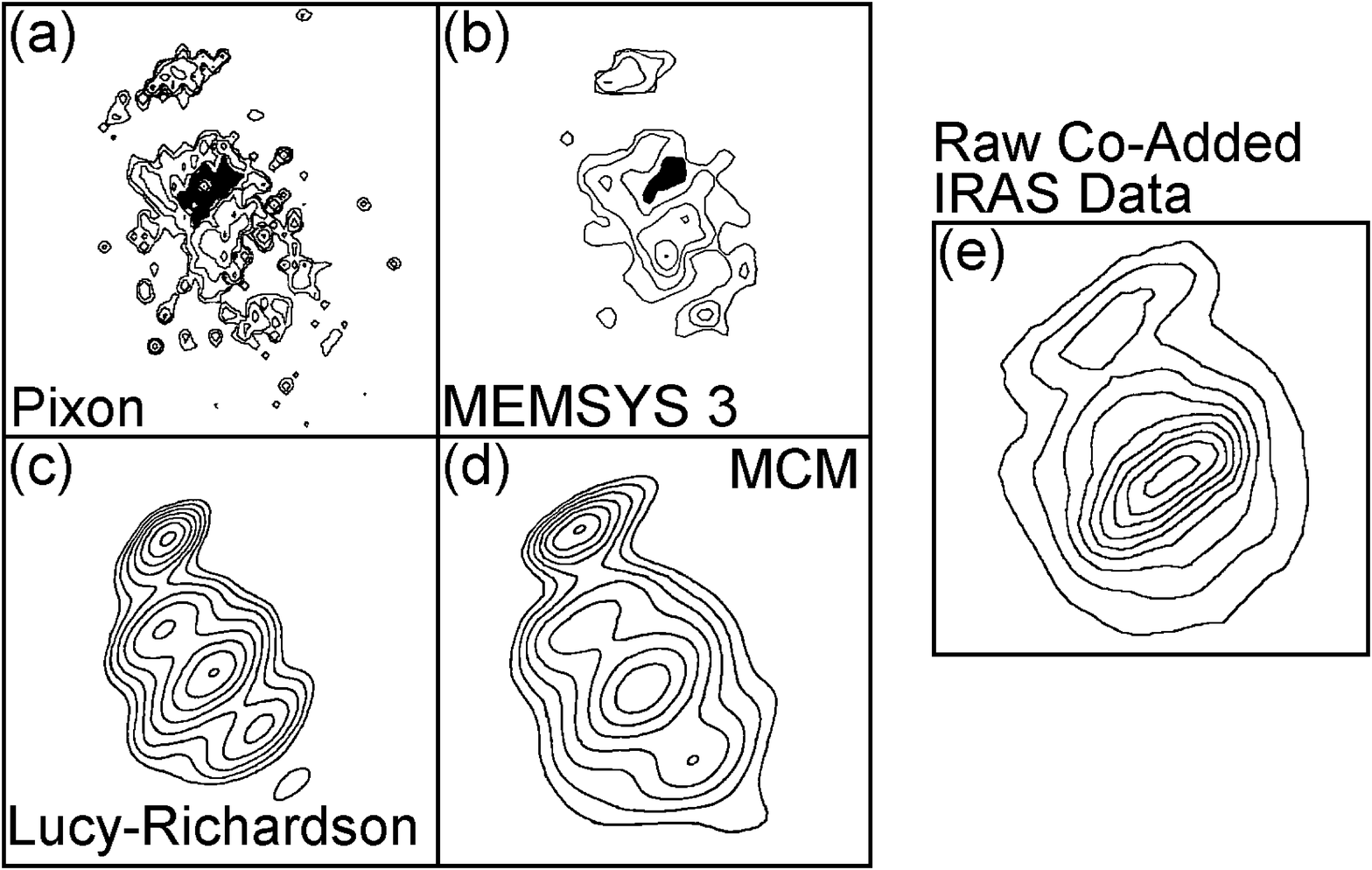}
\caption{Comparison of reconstruction methods for the interacting galaxy pair
M51: (a) Pixon method (Puetter \& Pi\~na 1994), (b) Maximum entropy method
(Bontekoe et al.\ 1991), (c) Lucy-Richardson method (Rice 1993), (d)
Maximum-Correlation Method, (Rice 1993, used by the HIRES package of NASA's
Infrared Processing and Analysis Center), and (e) the raw co-added
image.\label{fig:M51-comparison}}
\end{figure}

\subsection{IRAS Imaging of M51}

In Figure \ref{fig:M51-comparison} we present comparisons of reconstructed
images from 60$\mu m$ IRAS scans of the interacting galaxy pair M51.  This data
set was used in an image reconstruction contest at the 1990 MaxEnt Workshop
(Bontekoe 1991).  As can be seen, the Lucy-Richardson and maximum-correlation
(also known as HIRES) reconstructions fail to reduce image spread in the
cross-scan direction, i.e., the rectangular signature of the $1\farcm5\times
4\farcm75$ detectors is still clearly evident.  They also do not reconstruct
even gross features such as the ``hole'' (black region) in the emission north
of the nucleus---this hole is clearly evident in optical images of M51.  The
MEMSYS 3 reconstruction is significantly better, recovers the hole, and
resolves the NE and SW arms of the galaxy into discrete sources.  However, the
Pixon result is clearly superior.  In fact, its sensitivity is a factor of 200
higher than that of MEMSYS 3, and its linear spatial resolution is improved by
a factor of 3.  The reality of its minute details can be verified by comparing
with images at \htmladdnormallink{other wavelengths}
{http://www.pixon.com/figures/FIG_M51_recon.html}.

\begin{figure}
\plotone{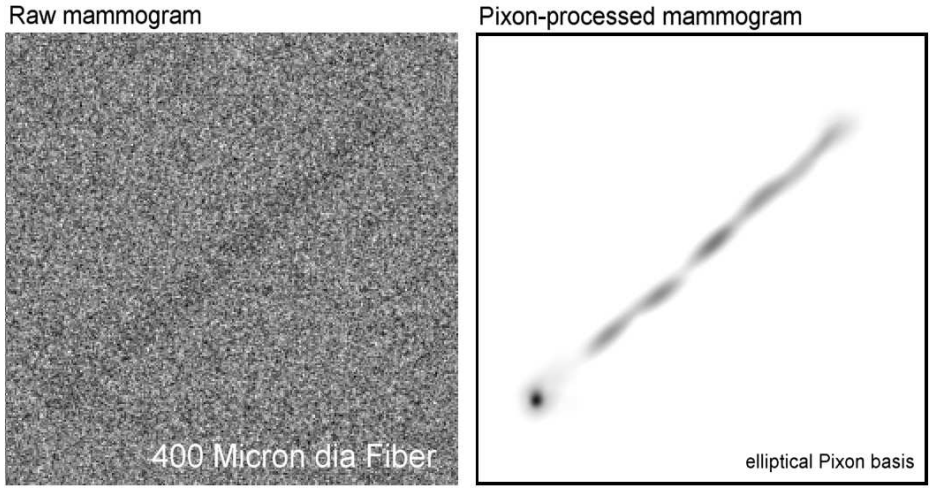}
\caption{Pixon reconstruction of X-ray mammography data.
\label{fig:mammogram}}
\end{figure}

\subsection{X-Ray Mammography}

Presented in Figure \ref{fig:mammogram} is a Pixon reconstruction of a standard
phantom from the American College of Radiology.  Here a fiber with a 400 micron
diameter was embedded in a piece of material with X-ray absorption properties
similar to the human breast.  This particular example was selected since it has
a low SNR.  A family of elliptical Pixon kernel functions was used for the
reconstruction.  The Pixon method easily detects the presence of the fiber.  In
some locations, however, the SNR of the fiber is so poor that statistically
significant signal is absent.  Hence, no detection can be made by the Pixon
method.

\acknowledgements

This work was was supported in part by NASA grants NAG53944 (to RCP) and
AR-07551.01-96A (to AY).


\begin{references}

\reference Bontekoe, T. R., 1991, in Maximum Entropy and Bayesian Methods,
eds.\ W. T. Grady Jr.\ \& L. H. Schick, (Dordrecht: Kluwer Academic
Publishers), 319

\reference Bontekoe, T. R., Kester, D. J. M., Price, S. D., de Jonge, A. R. W.,
\& Wesselieus, P. R., 1991, \aap, 248, 328

\reference Burch, S. F., Gull, S. F., \& Skilling, J., 1983, Comp.\ Vis.,
Graphics, \& Im.\ Process., 23, 113.

\reference Chaitin, G. J., 1966, J. Ass.\ Comput.\ Mach., 13, 547

\reference Donoho, D. L., 1995, IEEE Trans.\ Inf.\ Theory, 41, 613 $|$
\htmladdnormallink{Stanford report}
{http://www-stat.stanford.edu/reports/donoho/}
                                    
\reference Donoho, D. L. \& Johnstone, I., M., 1994, Comptes Rendus de
L'Acadamie Des Sciences Serie I-Mathematique, 319, 1317 $|$
\htmladdnormallink{Stanford report}
{http://www-stat.stanford.edu/reports/donoho/}

\reference Gull, S. F., 1989, in Maximum Entropy and Bayesian Methods, ed.\
J. Skilling, (Dordrecht: Kluwer Academic Publishers), 53

\reference Gull, S. F., \& Skilling, J., 1991, MemSys5 Quantified Maximum
Entropy User's Manual

\reference Kolmogorov, A. N., 1965, Inf.\ Transmission, 1, 3

\reference Pi\~na, R. K. \& Puetter, R. C.,\ 1993, \pasp, 105, 630

\reference Press, W. H., Teukolsky, S. A., Vetterling, W. Y., \& Flannery,
B. P, 1992, Numerical Recipes in Fortran, Second Edition, (Cambridge: Cambridge
University Press)

\reference Puetter, R. C., 1994, Proc.\ S.P.I.E., 2302, 112

\reference Puetter, R. C., 1995, Int.\ J. Image Sys.\ \& Tech., 6, 314

\reference Puetter, R. C., 1997, in Instrumentation for Large Telescopes, eds.\
J.\ M.\ Rodriguez Espinosa, A. Herrero, \& F. Sanchez, (Cambridge: Cambridge
University Press), 75

\reference Puetter, R. C. \& Pi\~na, R. K., 1993, Proc.\ S.P.I.E., 1946, 405

\reference Puetter, R. C., \& Pi\~na, R. K., 1994, in Science with high Spatial
Resolution Far-Infrared Data, (Pasadena: Jet Propulsion Laboratory), 95-4, 61

\reference Rice, W., 1993, \aj, 105, 67

\reference Shannon, C. E., 1948, in Key Papers in the Development of
Information theory, ed.\ D. Slepian, D., 1974, (New York, IEEE Press)

\reference Skilling, J., 1989, in Maximum Entropy and Bayesian Methods, ed.\
J. Skilling, (Dordrecht: Kluwer Academic Publishers), 45

\reference Solomonoff, R., 1964, Inf.\ Control, 7, 1

\reference Weir, N., 1991, in Proc.\ of the ESO/ST-ECF Data Analysis Workshop,
eds.\ P. Grosbo \& R. H. Warmels, 115

\reference Wiener, N. 1949, Extrapolation and Smoothing of Stationary Time
Series (New York: Wiley)

\end{references}
\end{document}